\documentstyle[psfrag,preprint,graphicx,tighten,eqsecnum,floats,aps]{revtex}

\begin{document}
\def\ni{\noindent}

\def\be{\begin{equation}}
\def\ee{\end{equation}}
\def\ba{\begin{eqnarray}}
\def\ea{\end{eqnarray}}
\def\b{$\bullet$}

\def\A{{\cal A}}
\def\H{{\cal H}}
\def\G{{\cal G}}
\def\N{{\cal N}}
\def\R{{\cal R}}
\def\Gauss{{\cal C}_{\rm Gauss}}
\def\C{{\cal C}}
\def\S{{\cal S}}
\def\I{{\cal I}}
\def\F{{\cal F}}
\def\HP{\H_{\rm poly}}
\def\HF{\H_{\rm Fock}}

\def\Gb{{\bar \G}}
\def\Ab{{\bar \A}}
\def\ab{{\bar A}}
\def\bg{{\bar g}}

\def\h{{\hat h}}
\def\hE{{\hat E}}
\def\hJ{{\hat J}}
\def\hA{{\hat A}}
\def\hC{{\hat \C}}
\def\hH{{\hat H}}
\def\hJ{{\hat J}}
\def\hK{{\hat K}}
\def\hL{{\hat L}}
\def\hR{{\hat R}}
\def\hT{{\hat T}}
\def\hq{{\hat q}}
\def\hV{{\hat V}}
\def\hG{{\hat \Gauss}}

\def\tr{{\rm Tr}}
\def\div{\rm div}
\def\const{{\rm const}}
\def\Cyl{{\rm Cyl}}
\def\cyl{\Cyl}
\def\su{{\rm su}}
\def\SU{{\rm SU}}
\def\diff{{\rm diff}}
\def\Diff{\C_{\rm Diff}}
\def\TDiff{{\rm TDiff}}
\def\hDiff{\hC_{\rm diff}}
\def\inv{{\rm inv}}
\def\phys{{\rm phys}}
\def\Tr{{\rm Tr\,}}

\def\dual{{\Cyl^P}^*}
\def\INV{{{\Cyl^P}^*}_\inv}
\def\a{V}
\def\g{\gamma}
\def\e{\epsilon}
\def\k{\kappa}
\def\t{\triangle}
\def\lp{{\ell}_{\rm Pl}}
\def\Comp{\bf C}
\def\Nat{\bf N}
\def\Real{\bf R}
\def\Rb{\bar{R}_{\rm Bohr}}
\def\r{{(r)}}

\def\smallr{\bf r}

\title{Polymer and Fock representations for a Scalar field}
\author{Abhay\ Ashtekar${}^{1,3}$, Jerzy\
Lewandowski${}^{2,1,3,4}$, and Hanno\ Sahlmann${}^{1,3}$}
\address{1. Center for Gravitational Physics and Geometry,
Physics Department, 104 Davey, Penn State, University
Park, PA 16802, USA\\
2. Institute of Theoretical Physics, University of Warsaw,
ul. Ho\.{z}a 69, 00-681 Warsaw, Poland\\
3. Erwin Schr\"odinger Institute, 9 Boltzmanngasse, 1090 Vienna,
Austria\\
4. Max Planck Institut f\"ur Gravitationsphysik, Albert Einstein
Institut, 14476 Golm, Germany}

\preprint{ESI-1233, CGPG-02/11-1}

\maketitle

\begin{abstract}

In loop quantum gravity, matter fields can have support only on
the `polymer-like' excitations of quantum geometry, and their
algebras of observables and Hilbert spaces of states can not refer
to a classical, background geometry. Therefore, to adequately
handle the matter sector, one has to address two issues already at
the kinematic level. First, one has to construct the appropriate
background independent operator algebras and Hilbert spaces.
Second, to make contact with low energy physics, one has to relate
this `polymer description' of matter fields to the standard Fock
description in Minkowski space. While this task has been completed
for gauge fields, important gaps remained in the treatment of
scalar fields. The purpose of this letter is to fill these gaps.

\end{abstract}

\section{Introduction}
\label{s1}

In this letter, we construct the `polymer representation' of a
real-valued scalar field (such as Klein Gordon) and compare it to
the Fock representation in Minkowski space-time. This work is
motivated by three considerations.

First, scalar fields provide a simple arena for mathematical
investigations of quantum fields in flat and curved space-times.
In loop quantum gravity, on the other hand, it is the gauge fields
that can be most easily incorporated because of the availability
of Wilson loops. For a Maxwell field, for example, not only has
the polymer representation been constructed in detail but its
relation to the Fock representation is also well-understood
\cite{mv,al7}. To make contact with the rich mathematical quantum
field theory literature, it is important to extend these results
to the case of a scalar field.

The second motivation comes from the fact that the `polymer'
description of a real-valued scalar field and, in particular, the
notion of shadow states used in the semi-classical analysis, is
technically more subtle than that in the Maxwell case. As we will
see below, there is a sense in which the gauge group $U(1)$ of the
Maxwell theory is now replaced by the additive group of real
numbers. Therefore, at the technical level, this analysis provides
a first step for incorporation of non-compact gauge groups in the
polymer representation.

The third motivation comes from loop quantum gravity itself. In
the discussion of the Hamiltonian constraint either in full loop
quantum gravity \cite{tt} or in the more restricted context of
loop quantum cosmology \cite{mb}, the detailed treatments have
focussed only on the $\sigma$-model type scalar fields which take
values in compact groups. Incorporation of Klein-Gordon type
scalar fields requires an extension of the kinematical framework.
Conceptually, the existence of such an extension is also needed in
the discussion of black hole entropy in quantum geometry,
particularly in the proof that the Hawking-Bekenstein formula is
correctly \cite{iw} modified in presence of a non-minimally
coupled scalar field \cite{ac}.

The letter is organized as follows. In part II, we present the
`polymer description' of a scalar field (see also \cite{okolow}).
Since we wish to compare it with the Fock representation, we will
consider scalar fields in Minkowski space. However, this
discussion can be trivially generalized to any manifold with
topology $\Sigma \times R$ where $\Sigma$ is any spatial
3-manifold. In part III we recast the Fock representation in a
form which facilitates comparison with the polymer description.
This construction parallels that in the Maxwell theory \cite{al7}
which in turn was inspired by key ideas introduced by M.
Varadarajan \cite{mv}. The comparison is carried out in part IV.
Part V presents the overall viewpoint from a quantum geometry
perspective.

\section{The polymer quantum scalar field}
\label{s2}

As in quantum geometry \cite{qg}, we will proceed in the following steps: i)
Select an algebra of functions on the classical configuration
space which are to serve as `elementary' classical observables;
ii) Select a suitable representation of this algebra by operators
on a Hilbert space; and, iii) Express momenta as self-adjoint
operators on this Hilbert space. Our choices will ensure that the
polymer scalar field can `live' on quantum geometry.

For scalar fields of interest, the classical configuration space
$\A$ consists of the space of all real valued, smooth functions
$\phi$ (with an appropriate fall-off) on the spatial 3-plane,
$R^3$. We will first introduce `holonomy' functions on $\A$ which
will constitute the \emph{elementary configuration variables} for
the polymer representation. Let us begin with some definitions. We
will follow the terminology used in the polymer framework for
Maxwell (or, more generally, gauge) fields. A set $\a$ consisting
of a finite number of points on $R^3$ will be called a
\emph{vertex set}. (The empty set is allowed as a vertex set.)
Given a vertex set $\a = \{x_1, \ldots x_n \}$, denote by
$\cyl_\a$ the vector space generated by finite linear combinations
of the following functions of fields $\phi$
\be \N_{\a,{\vec \lambda}}(\phi) = e^{i\sum_j \lambda_j\,
\phi(x_j)}\ee
where $\vec{\lambda} \equiv (\lambda_1, \lambda_2, \ldots
\lambda_n)$ are arbitrary real numbers. $\cyl_\a$ has the
structure of a $\star$ algebra. Finally we introduce the space
$\cyl$ of \emph{all} cylindrical functions on $\A$:
$$ \cyl = \cup_\a \, \cyl_\a$$
We can complete it with respect to the sup norm and obtain a
$C^\star$ algebra which we denote by $\overline{\cyl}$. This can
be taken to be \emph{the $C^\star$ algebra of configuration
observables.}

Our next task is to find a suitable representation of this
algebra. Since $\overline{\cyl}$ is an Abelian $C^\star$-algebra
with identity, we can use the Gel'fand theory to conclude that
every of its cyclic representations is of the following type: The
Hilbert space $\H$ is the space $L^2(\Ab, d\mu)$ of square
integrable functions on a compact topological space $\Ab$ with
respect to some regular Borel measure $\mu$, and $\overline\cyl$
acts on $\H$ by multiplication. $\Ab$ is called
the Gel'fand spectrum of the algebra $\overline\cyl$.%
\footnote{In what follows only the following facts about $\Ab$ are
important: i) $\Ab$ is a compact topological space, the
$C^\star$-algebra of \emph{all} continuous functions on which is
naturally isomorphic with $\overline{\cyl}$. (For notational
simplicity, we will identify the two $C^\star$ algebras); and, ii)
There is a natural, dense embedding of the classical configuration
space $\A$ into $\Ab$.}
Because $\H = L^2(\Ab, d\mu)$, it is natural to think of $\Ab$ as
the `quantum configuration space'.

To specify the structure of $\Ab$, let us first consider a vertex
set $\a_o$ consisting of a single point $x_o$. Then, $\cyl_{\a_o}$
is the space of all almost periodic functions on a real line
(given by equivalence classes of $\phi(x)$ where two are
equivalent if they have the same value at $x_o$). The Gel'fand
spectrum of the corresponding $C^\star$-algebra
$\overline{\cyl}_{\a_o}$ is the Bohr completion $\Rb^{x_o}$ of the
real line $R$. That is, $\Rb^{x_o}$ is a compact topological space
such that $\overline{\cyl}_{\a_o}$ is the $C^\star$ algebra of
\emph{all} continuous functions on $\Rb^{x_o}$. $R$ is densely
embedded in $\Rb^{x_o}$; thus $\Rb^{x_o}$ can be regarded as a
completion of $R$. At the point $x_o$, whereas the classical
fields take values $\phi(x_o)$ in $R$, the quantum field will take
values in $\Rb^{x_o}$. Thus, the classical configuration space
$\A$ is now enlarged to a quantum configuration space $\Ab$
consisting of (arbitrarily discontinuous) $\Rb$-valued functions
on $R^3$.

\emph{Remark}: In the existing treatments of scalar fields in the
polymer representation, the $\lambda_j$ are generally restricted
to be integers. Then, the configuration variables are
\emph{periodic} functions of $\phi(x_o)$ and the Gel'fand spectrum
of $\overline{\cyl}_{\a_o}$ is $S^1$. Physically, this corresponds
to considering a $U(1)$ $\sigma$-model in Minkowski space where
the fields are required to take values in $U(1)$
---rather than $R$--- at each point of the spatial plane $R^3$.
For the Klein-Gordon type real scalar fields, on the other hand,
periodic functions do not suffice to separate points of the
configuration space $\Ab$. Almost periodic functions are essential
and this in turn leads to the Bohr compactification. (For more
information on almost periodic functions and the Bohr
compactification see for example \cite{bohr}.)

The polymer representation is based on a \emph{preferred,
diffeomorphism invariant, faithful Borel measure $\mu_o$ on
$\Ab$}. This measure is defined by the positive linear functional
$\Gamma_o$ on $\overline\cyl$:
$$\Gamma_o (\N_{\a, \vec{\lambda}}) = \cases{ 1 & if $\lambda_j
=0, \, \forall j$\cr 0 & otherwise\cr}$$
Since continuous functions on $\Ab$ are given precisely by the
(Gel'fand transform of) elements of $\overline{\cyl}$, to define a
regular Borel measure, it suffices to specify values of integrals
of all these functions. The measure $\mu_o$ defined by $\Gamma_o$
is given by:
\be \int_{\Ab}  d\mu_o \,\, (\N_{\a, \vec{\lambda}}) = \cases{ 1 &
if $\lambda_j =0, \, \forall j$\cr 0 & otherwise\cr}\ee
The quantum Hilbert space of the polymer representation is $\HP =
L^2(\Ab, d\mu_o)$. With each pair $(x_o,\lambda_o)$, there is an
`elementary' configuration/holonomy operator $\hat{h}(x, \lambda)$
which acts by multiplication: For all cylindrical functions
$\Psi(\phi)$ we have
\be \label{hol1} \hat{h}(x_o, \lambda_o)\, \Psi(\phi) =
e^{i\lambda_o \phi(x_o)}\, \Psi (\phi) \ee
These operators are unitary. But because the positive linear
functional $\Gamma$ is discontinuous in $\lambda$, $h(x,\lambda)$
fail to be weakly continuous in $\lambda$ whence there is no
operator $\hat \phi(x)$ on $\HP$. This is completely analogous to what
happens in the polymer representation of the Maxwell field where
the holonomy operators $\hat{h}_e$ are well defined for every edge
$e$ but the connection operator $\hat{A}$ is not.

Finally, we turn to the momentum operators. For the classical
scalar field, with each test function $g$ on $R^3$ one associates
a momentum functional
$$
P(g) = \int_{R^3}d^3x\,\, \pi g
$$
where $\pi$ is the momentum conjugate to $\phi$. Note that this
definition does not require the volume form on $R^3$ since $\pi$
by itself is a density of weight one. The Poisson brackets between
these momenta and the holonomy functionals are given by:
\be \label{poisson} \left\{P(g),h(x_o,\lambda_o)\right\}=
i\lambda_o\,g(x_o)h(x_o,\lambda_o) \ee
To implement these relations in the quantum theory, for each test
function $g$ on $R^3$ we define a momentum operator $\hat{P}(g)$
whose action on $\cyl$ is given by:
\be \label{mom1} [\hat{P}(g) \, \N_{\a, \vec{\lambda}}](\phi) =
\left( \hbar \, \sum_j \lambda_j g(x_j)\right) \, \, \N_{\a,
\vec{\lambda}}(\phi)  \ee
These operators are essentially self-adjoint on $\HP$. They are
the analogs of the smeared electric field operators in the Maxwell
case. Their eigenvectors are simply $\N_{\a, {\vec{\lambda}}}$. In
the Maxwell case, since the gauge group is $U(1)$ rather than $R$,
in place of the real numbers $\lambda_j$, we had
\emph{integers} $n_j$; the present $\lambda_j$ are the continuous analogs of
`fluxons' (or `spins' of quantum geometry). As in quantum Maxwell
theory or quantum geometry, we can decompose $\HP$ as a direct sum
of Hilbert spaces associated with vertex sets:
$$ \HP = \bigoplus_{\a} \H_{\a} $$
where the direct sum is over arbitrary vertex sets $\a\equiv\{x_1,
\ldots x_n\}$ (including the empty vertex set) and each $\H_\a$ is
the Cauchy completion of finite linear combinations of $\N_{\a,
{\vec \lambda}}$, where each $\lambda_j$ is \emph{non-zero}. (If a
$\lambda_j$ vanishes, that function does not depend on $\phi(x_j)$
and thus appears in the Hilbert space $\H'_{\beta}$ associated
with a vertex set $\beta$ with one less point, $x_j$.) Thus, an
orthonormal basis in $\HP$ is given by the functions $\N_{\a,
{\vec \lambda}}(\phi)$, where ${\vec \lambda} \equiv
\{\lambda_1,\ldots \lambda_N \}$ are \emph{non-zero} real numbers.
Following the terminology in quantum geometry, we will refer to
these basis vectors as \emph{scalar network functions}.

Next, let us note the commutation relations between the `holonomy'
and the momentum operators in the polymer representation. The
`holonomy' operators commute among themselves and so do the
momentum operators. The only non-trivial commutators are
\be \label{ccr1} [\hat{P}(g), \, \hat{h}(x_o, \lambda_o)]\, = \,
\hbar \lambda_o \, g(x_o)\,\, \hat{h}(x_o, \lambda_o)\ee
These commutation relations implement the Poisson relations
(\ref{poisson}) among the
classical variables and will be used in section \ref{s3}. Note
that the representation of the holonomy-momentum algebra we have thus
obtained is \emph{irreducible} and was constructed without \emph{any}
reference to a background field such as a metric; the measure
$\mu_o$ is diffeomorphically invariant and all our constructions
are diffeomorphically covariant.

\emph{Remark}: While we worked in the continuum to facilitate
comparison with the Fock representation, our construction is
motivated by the requirement that the polymer description of the
scalar field be well-defined also on a quantum geometry \cite{qg}.
In this case, the fundamental excitations of geometry are
1-dimensional, polymer-like, and $R^3$ is replaced by arbitrary
graphs (which can be regarded as `floating lattices'). Geometry
and gauge fields have support on these graphs. Scalar fields on
the other hand will reside only at vertices. Thus, by identifying
the `vertex-sets' of this section with the set of vertices of the
standard graphs of quantum geometry, we can do physics of the
quantum scalar field on quantum geometry \cite{st}.

\section{The $\smallr$-Fock description}
\label{s3}

The standard Fock description can be cast in the following
convenient form. The Hilbert space $\HF$ is $L^2(\S^\star,
d\mu_F)$, where $\S^\star$ the space of tempered distributions on
$R^3$ and $d\mu_F$ is the Gaussian Fock measure on it. The basic
operators can be taken to be $\hat{U}(f) =\exp i \int d^3x
\,\hat{\phi}(x) f(x)$ and $\hat{\pi} (g) = \int
d^3x\,\hat{\pi}(x)g(x)$, where $f$ and $g$ are test functions on
$R^3$. The first act by multiplication:
\be [\hat{U}(f)\, \Psi ](\tilde\phi)\, = \, e^{i\int d^3x
\tilde\phi(x) f(x)}\,\,\,  \Psi(\tilde\phi)\, , \ee
where $\tilde\phi \in \S^\star$. These operators are unitary, and
$\hat{U}(\lambda f)$ are weakly continuous in the real parameter
$\lambda$, whence $\int d^3x  \hat\phi (x) f(x)$ exist as
self-adjoint operators. The momentum operators act via derivation:
\be [\hat{\pi}(g)\, \Psi](\tilde\phi) \, = \, \frac{\hbar}{i}
\big[\,\int d^3x\, g(x)\, \frac{\delta}{\delta \tilde\phi(x)}\,
-\, \int d^3x\, g(x)\, \Delta^{\frac{1}{2}}\, \tilde\phi(x)\,
\big]\, \Psi (\tilde\phi) \ee
where the second term arises because the `divergence' of the
Gaussian Fock measure with respect to the vector field
$\int d^3x\, g(x) \delta/\delta\phi(x)$ is non-zero.

We will now construct an isomorphic description in which Fock
states are represented as \emph{square-integrable functions on
$\Ab$ with respect to a new measure} $\mu_F^\r$ and discuss the
action of operators. This step will enable us to regard both $\HP$
and $\HF^\r$ as Cauchy completions of $\cyl$ (with respect to
$\mu_o$ and $\mu_F^\r$). Since states in both representations
arise as functions on the same space $\Ab$, it will be easier to
compare them.

First, for each real number $r>0$, we define a `taming map'
$\Lambda_r$ from $\S^\star$ to $\Ab$ as follows. Fix a 2-point
smoothening function $f_r(x,y)$ on $R^3$ such that: i) $f_r(x,y)$
is symmetric in $x,y$, ii) $f_r(x,y) = g_r(x-y)$ where $g_r$ is a
Schwartz test function; and, iii) as a distribution, $f_r(x,y)$
tends to $\delta (x,y)$ as $r$ tends to zero. A concrete example
is provided by
\be f_r (x,y)= \frac{1}{(2\pi)^{\frac{3}{2}}}\,
\frac{e^{-\frac{|x-y|^2}{2r^2}}}{r^3}\ee
We will often fix $x$, regard $f_r$ as a function of $y$ and write
$f_r(x,y) =f_{r, x}(y)$. Now, given any tempered distribution
$\tilde{\phi}$, we will set:
\be [\Lambda_\r\, (\tilde{\phi})](x) = \int d^3y \, f_{r,x}(y)
\tilde{\phi}(y)\, .\ee
The result, $\Lambda_\r\, (\tilde{\phi})$, is a $C^\infty$
function on $R^3$ and, in particular, defines an element of $\Ab$.
It is easy to verify that the linear map,
$$\Lambda_\r: \, \S^\star \mapsto \Ab \, ,$$
is injective. Denote by $\mu_F^\r$ the push-forward of the Fock
measure on $\S^\star$. Then, each element of the Fock space can
also be represented as a function on $\Ab$ which is square
integrable with respect to $\mu_F^\r$. Thus, the Fock space $\HF$
is naturally isomorphic with $\HF^\r = L^2(\Ab, d\mu_F^\r)$. We
will call it the r-Fock space. Let us explore the new measure on
$\Ab$. Since $\overline{\cyl}$ is the space of \emph{all}
continuous functions on $\Ab$, any regular Borel measure is
determined by specifying the integrals of functions in $\cyl$. By
linearity, it suffices to calculate the integrals of the
`elementary cylindrical functions' $\N_{\a, \vec{\lambda}}
(\phi)$. We have:
\be \int_\Ab d\mu_F^\r\,\, \N_{\a, \vec{\lambda}}(\phi)\,
 = \,\int_{\S^\star} d\mu_F \,\, e^{i\int d^3y\, \big(\sum_j \lambda_j
f_{r,x_j}(y)\big)\,  \tilde\phi(y)} = e^{-\frac{1}{4}|| \sum_j
\lambda_j f_{r,x_j}(y)||^2} \ee
for all $\a$ and $\vec{\lambda}$, where
$$ ||F||^2 = \int d^3y\, F(y) \Delta^{-\frac{1}{2}}\, F(y).$$
Here, in the last step we have used the well known fact about the
standard Fock measure. Thus, we have exhibited the Fourier
transform of the measure $d\mu_F^\r$, which characterizes it
completely. In particular, each cylindrical function is
normalizable and thus determines an element of the r-Fock space.
It is interesting ---and perhaps even counter-intuitive-- that the
r-Fock space and the polymer Hilbert space can both be obtained by
Cauchy completing $\cyl$ with suitable measures, namely,
$d\mu_F^\r$ and $d\mu_o$ respectively.

Next, let us consider the r-images of the basic operators
$\hat{U}(f)$ and $\hat\pi(g)$. Let us first consider the image of
$\hat{U}(\lambda f_{r,x})$ where we have restricted the smearing
function $f$ to have the form $f = \lambda f_{r,x}$ for some
$\lambda \in R$ and $x\in R^3$. A straightforward calculation then
yields the following action of the r-image of this operator on
cylindrical functions $\Psi(\phi)$:
\be \label{hol2} [\hat{U}_\r(\lambda_o f_{r,x_o})\, \Psi](\phi)  =
e^{i\lambda_o \phi(x_o)}\, \Psi(\phi)\ee
Thus, as one might expect, the action is just by multiplication.
However, it is interesting that the result is again a cylindrical
function and, on the right side, there is no reference to the
detailed form of the taming function; only $\lambda_o$ and $x_o$
appear. Note that since the operator $\hat{U}(\lambda f)$ is
weakly continuous in $\lambda$ on $\HF$, the operator $\hat{U}_\r
(\lambda f_{r,x_o})$ is weakly continuous in $\lambda$ on
$\HF^\r$. Hence, although multiplication by $\phi(x_o)$ does not
leave $\cyl$ invariant, it is a well-defined operation on
$\HF^\r$. (Recall this was not the case on $\HP$.) Finally, we
have given the explicit formula only when the test functions are
of the form $\lambda f_{r,x}$. However, since the vector space
generated by these test functions (with arbitrary $\lambda \in R$
and $x \in R^3$) is dense in the space of all test functions, this
specification suffices.

Next, let us specify the action of the image of the Fock momentum
operators on cylindrical functions. Again, we can first restrict
ourselves to test functions of the form $g = \lambda f_{r,x}$ and
then extend the action of the operator to arbitrary test functions
$g$. We have:
\be [\hat\pi_{(r)}(\lambda_o f_{r,x_o}) {\cal N}_{V,
\vec\lambda}](\phi) = \hbar [ \int d^3y \lambda_o f_{r,x_o}(y)
\sum_j \lambda_j f_{r,x_j}(y) + i \lambda_o
(\Delta^{\frac{1}{2}}\, \phi)(x_o) ] {\cal N}_{V,
\vec\lambda}(\phi)\ee
(Note that $\Delta^{\frac{1}{2}} \phi$ is well-defined for all
$\phi$ in the image of the `taming map' $\Lambda_{(r)}$, i.e., on
the support of the measure $\mu_F^\r$.) Again the second term is
the `divergence' of the vector field `$g (\delta/\delta
\tilde{\phi})$' with respect to the measure $\mu_F^\r$. Finally,
the only non-trivial commutators between the holonomy and momentum
operators are:
\be \label{ccr2} [\hat{\pi}_\r(g), \, \hat{U}_\r (\lambda_o
f_{r,x_o})] = \hbar \lambda_o \, [\int d^3y\, g(y) f_{r,x_o}(y)]\,\,
\hat{U}_\r (\lambda_o f_{r,x_o}) \ee

We conclude with two conceptually important remarks.

i) The Fock and the r-Fock representations are naturally
\emph{isomorphic}. Therefore, everything one can do in the
standard Fock representation ---e.g., the introduction of the
notion of Hadamard states and regularization of products of
operators--- can be translated \emph{unambiguously} to the r-Fock
representation, i.e. to structures defined on $\Ab$. However, as
with the action of the momentum operators, these constructions can
involve the operation of multiplication by $\phi(x)$ which fails
to be well-defined in the polymer Hilbert space $\HP$.

ii) From the strict perspective of Minkowskian field theories, the
presence of the taming function $f_{r,x}(y)$ (and especially its
new scale $r$) in all these expressions seems awkward. From the
quantum geometry perspective, this is because the `physical
origin' of this function lies beyond the scope of the continuum
theories. The viewpoint is that, in Nature, there is fundamental
discreteness because of the quantum nature of geometry and the
continuum geometry appears only on suitable coarse graining of the
fundamental, quantum geometry. It is this coarse graining that
provides the taming function. In effect, the discrete geometry can
be approximated by the continuum only if we define `effective
geometric fields' at each point $x$ in the continuum by averaging,
i.e., by smearing  the fundamental quantum-geometric structures
with a function $f_{r,x}(y)$. On general grounds, one expects the
discreteness to appear at the Planck scale $\ell_P$. Hence the
averaging length scale $r$ has to be much larger than $\ell_P$ but
much smaller than the length scales probed by the energies
available in given experiments. Thus, as one `descends from the
fundamental, Planck scale perspective', one will be led to an
effective description involving $f_{r,x}$ ---as manifested, e.g.,
in the r-Fock representation. But one can then notice that by
`undoing the taming map' one can get rid of $f_{r,x}$ provided one
works with the space $\S^\star$ of distributions. From this
perspective, then, r-Fock representation is an intermediate step;
it is awkward because it has neither the mathematical elegance
that working in the continuum from the beginning brings nor the
deeper physical insights that working with quantum geometry can
bring.

\section{Comparison}
\label{s4}

Let us begin with algebras. The standard Weyl algebra can be
obtained by a suitable completion of the algebra generated by the
operators $\hat{U}(\lambda f_{r,x})$ and $\hat{V}(g) := \exp i
\hat{\pi}(g)$. These operators satisfy the well-known commutation
relations
\be \hat{V}(g)\, \hat{U}(\lambda f_{r,x}) = e^{i\hbar\lambda \int
d^3y f_{r,x}(y) g(y)}\,\, \hat{U}(\lambda f_{r,x})\,\hat{V}(g).
\ee
On the polymer side, consider operators $\hat{h}(x,\lambda)$ and
$\hat{{\cal V}}(f_{r} \star g) = \exp i \hat{P}(f_{r,x} \star g)$,
where $(f_{r}\star g)(y) = \int d^3x f_r(x,y) g(x)$ is the
convolution of $f_r$ and $g$. Their commutation relations are
\be {\hat{{\cal V}}}(f_{r}\star g) \hat{h}(x, \lambda) =
e^{i\hbar\lambda \int d^3y f_{r,x}(y) g(y)}\,\, \hat{h}(x,
\lambda) {\hat{{\cal V}}}(f_{r} \star g) \ee
Hence,
$$(\hat{U}(\lambda f_{r,x}), \, \hat{V}(g)) \longmapsto
(\hat{h}(x,\lambda),\, \hat{{\cal V}}(f_{r}\star g))$$
defines a *-isomorphism between the two algebras. Put differently, in
this letter we have discussed two inequivalent representations of
the standard Weyl algebra; the Fock and the Polymer.

Next, let us now compare the polymer and the r-Fock
representations. Both Hilbert spaces, $\HP$ and $\HF^\r$, can be
obtained by Cauchy completing $\cyl$ but using inner products
determined by the respective measures $d\mu_o$ and $d\mu_F^\r$. On
$\cyl$, the holonomy operators of the polymer representation and
the configuration operators of the $r$-Fock representation are
related by (see (\ref{hol1}) and (\ref{hol2})):
$$ [(\hat{h}_{\lambda_o, x_o} - \hat{U}_\r (\lambda_o f_{r,x_o})]
  \, \Psi = 0$$
for all cylindrical functions $\Psi$ on $\Ab$. Next, let us
consider the momentum operators. In the polymer representation we
have (see (\ref{ccr1}))
$$[\hat{P}(g), \, \hat{h}(x_o, \lambda_o)]\, = \, \hbar \lambda_o
\, g(x_o)\, \hat{h}(x_o, \lambda_o) $$
while in the r-Fock representation we have (see (\ref{ccr2}))
$$[\hat\pi_\r (g), \,  \hat{U}_r (\lambda_o f_{r,x_o})] = \hbar
\lambda_o\,[\int\, d^3x g(x) f_{r,x_o}(x)]\, \hat{U}_r (\lambda_o
f_{r,x_o})]
$$
Therefore, the appropriate operators to compare are
$$\hat{P}(g\star f_{r}) \quad {\rm and}\quad  \hat{\pi}_\r (g) \, .$$
In the action of these operators on $\cyl$, the first term (`Lie
derivative along the vector fields') is the same. But while the
vector fields on $\Ab$ defining these momenta are divergence free
with respect to $\mu_o$, they are not with respect to $\mu_F^\r$.
Hence in the expression of $\hat{\pi}_r(x)$ there is an extra
term. It is this difference that makes the two representations of
the Weyl algebra unitarily inequivalent.

Finally, let us compare the two measures. For this, we can
calculate the integrals of general elements of $\cyl$ with respect
to both. Given a vertex set $\a = \{ x_1, \ldots x_n\}$, let us
define
$$G_{ij} = \int d^3y \,\, f_{r,x_i}(y)\, \Delta^{-\frac{1}{2}}
f_{r,x_j}(y)\, . $$
For basis functions $\N_{\a, {\vec{\lambda^o}}}(\phi)$ associated
with $\a$, we have:
\ba & &\int_\Ab d\mu_o\, [\sum_{\vec \lambda} \, e^{-\frac{1}{4}
\sum_{j,k}G_{jk}\lambda_j \lambda_k} \bar{\N}_{\a,
\vec{\lambda}}(\phi)]\,\, \N_{\a, {\vec{\lambda_o}}}(\phi)\nonumber\\
&=&  \,\, \sum_{\vec \lambda}\,\, e^{-\frac{1}{4}\sum_{j,k} G_{jk}
\lambda_j \lambda_k}\,\, \int_\Ab d\mu_o\,\,  \overline{\N}_{\a,
\vec{\lambda}}(\phi)\,\, \N_{\a,
{\vec{\lambda_o}}}(\phi)  \nonumber\\
&=& e^{-\frac{1}{4} \sum_{ij}\, G_{ij}\, \lambda^o_i\lambda^o_j}
\nonumber \\
&=& \int_\Ab d\mu_F^\r \, \, \N_{\a, {\vec{\lambda_o}}}(\phi) \ea
Therefore, we can write the relation between the two measures on
$\Ab$ as
\be\label{relation} d\mu_F^\r (\phi) = [\,\sum_{\a, {\vec
\lambda}}\, \, e^{-\frac{1}{4} \sum_{j,k}G_{jk} \lambda_j
\lambda_k} \overline{\N}_{\a, \vec{\lambda}}(\phi) \, ]\,\, d\mu_o
(\phi) \, . \ee
Since the sum is over \emph{continuous} variables $(\a, {\vec
\lambda})$, the quantity in square brackets is \emph{not} a
function on $\Ab$. In fact, using the results of \cite{jv}, one
can show that the two measures are inequivalent.%
\footnote{In particular, there is \emph{no} state in $\HP$ for
which the expectation value of the holonomy $\exp i[\lambda_o
\phi(x_o)]$ equals $\int_\Ab d\mu_F^\r \exp i[\lambda_o
\phi(x_o)]$ for all pairs $x_o,\lambda_o$.}
Thus, the equation is to be understood only in the sense of
distributions: every element of $\cyl$ is integrable with respect
to both sides and the value of the integral with respect to the
measure on the right side equals that on the left.

Let us summarize. The $C^\star$ algebra generated by
$\hat{U}(\lambda f_{r,x})$ and $\hat{V}(g)$ ---i.e., the standard
Weyl algebra ---admits two unitarily inequivalent representations.
In both representations a dense subset of states is provided by
$\cyl$; the quantum configuration space $\Ab$ provides a `common
home'. However, the completions are with respect to inequivalent
measures, $d\mu_o$ and $d\mu_F^\r$ on $\Ab$. The configuration
operators act by multiplication on $\cyl$ in both cases. But the
action of the momentum operators differs. In the polymer
representation, the unitary operators $\hat{U}(\lambda f_{r,x})$
fail to be weakly continuous in $\lambda$ whence $\hat\phi (x)$
fail to exist as operator-valued distributions while in the
$r$-Fock representations the weak continuity holds and the
operator-valued distributions exist. The polymer representation
makes no reference to a metric on $R^3$; all constructions are
covariant with respect to the diffeomorphism group on $R^3$. The
$r$-Fock representation, of course, is tied to the flat metric.

\emph{Remark}: In the Maxwell case, the $\lambda_j$ are replaced
by integers $n_j$ and the vertex sets $\a$ by graphs $\gamma$.
Therefore, when we restrict ourselves to a fixed graph
$\g$, in the analog of (\ref{relation})
the only sum involved is over integers. The restriction
$d\mu_{F, \g}^\r$ of the r-Fock measure and the restriction
$d\mu_{o,\g}$ of the polymer measure to the graph $\gamma$ are related by a
continuous function, whence these two measures are \emph{absolutely
continuous} with respect to one another \cite{al7}. In the scalar
field case, since $\lambda_j$ are continuous labels, even the
restrictions of the two measures to any one vertex set
$\a$ fail to be absolutely
continuous. To obtain absolute
continuity, one has to further restrict the $\lambda_j$s to a
\emph{countable} set. Consequently, the discussion of \emph{shadow
states} \cite{al7} now requires this additional restriction.

\section{Outlook}
\label{s5}

Introduction of the polymer representation raises two obvious
questions: i) We know that the Fock representation can be used
very effectively to describe low energy physics. How would this
description arise from the `fundamental' theory which is based on
quantum geometry and polymer description of quantum fields?; and,
ii) Can the `fundamental' framework address any of the open
problems of quantum field theory?

A comprehensive answer to these questions would require a step by
step procedure which starts from the solutions to the quantum
constraints in the coupled gravity and matter theory and analyzes
them in detail in the semi-classical sector of the theory. Such a
detailed reduction is yet to be constructed. So, we will adopt an
optimistic viewpoint, assume that the missing intermediate steps
can be filled, and summarize the final picture envisaged today.

The semi-classical state of quantum geometry corresponding to
Minkowski space will provide a graph $\g$ and a quantum geometry
state on it (more precisely \cite{al7}, an element of $\cyl^\star$
of quantum geometry which assigns amplitudes to each state defined
on a \emph{suitable} family of graphs $\g$). Roughly, the edge
lengths and the average separation between vertices of these
graphs would be a Planck length $\ell_P$ (as measured by the
continuum geometry at which the semi-classical state is peaked).
The flat continuum geometry would arise only when we coarse grain
the state with a smoothening function $f_{r,x}$ with $\ell_P\ll r
\ll \hbar/ E$ where $E$ is the highest energy scale we are
interested in. This smoothening function will be then used also in
constructing the r-Fock representations of matter fields. In what
follows, for simplicity we will refer to quantum field theories on
given quantum geometries as `fundamental' and regard the coarse
grained continuum description as an approximation.

Let us return to the scalar field. The `fundamental' scalar field
will reside only on the vertex set $\a$ of each of these graphs
$\gamma$. Thus, qualitatively, we have a lattice field theory. If
we only consider states in $\cyl_\a$, the restricted polymer and
r-Fock descriptions will be unitarily equivalent (once the
restriction mentioned in the remark at the end of section IV is
made). Basically, this is an illustration of Fell's theorem
\cite{fell}: from the continuum perspective we are looking at a
restricted class of observables in both theories.  But the
viewpoint is that \emph{this restricted framework is a more
fundamental description than the continuum one}. Recall that
predictions of renormalizable theories for energy scales $E$ are
insensitive to the details of the structure at length scales $L
\ll \hbar/E$. Therefore,  for such theories, the restricted
framework will provide a (generalized) lattice description whose
predictions are borne out in the low energy experiments.
Observationally, these predictions will be indistinguishable from
those of the continuum Fock theory. From the `fundamental' point
of view, these predictions are really derived from the polymer
description, restricted to the graphs selected by the
semi-classical state of the quantum geometry and the remark about
agreement with the r-Fock description is a only a quick way to
check that these results are observationally viable. Thus, the
viewpoint is that neither the r-Fock nor the polymer description
in the continuum is fundamental. However, the continuum Fock
description is a very useful approximation while the continuum
polymer scalar field description is not likely to be directly
useful in the familiar situations.%
\footnote{For \emph{gauge fields}, the continuum
polymer description  may be useful, e.g., in describing type II
superconductors where flux of the magnetic field is quantized; or,
in understanding some issues related to confinement in QCD.}
The real arena for the polymer description is quantum geometry.

The `fundamental' description should be able to shed new light on
quantum field theory issues. We will conclude with some examples.
First, all results of the `fundamental theory' are expected to be
finite since by construction the theory would be free of
ultraviolet divergences.%
\footnote{Note that, although many of the techniques used are the
same as in lattice gauge theory, the viewpoint here is
diametrically opposite. Here (generalized) lattices are not
convenient mathematical constructs to approximate the continuum
theories; they are provided by the physical quantum geometry and
therefore \emph{more} fundamental than the continuum. We can and
do work with the continuum limit but primarily for mathematical
convenience, e.g., because differential equations and integrals
are often easier to control than difference equations and discrete
sums.}
Therefore it should be possible to trace the precise manner in
which the continuum approximation leads to these divergences.
Second, it may provide a new physical basis, rooted in quantum
geometry, for the Wilsonian ideas of renormalization flows. As we
increase the energy scale $E$ (still keeping it well below the
Planck scale) we have to use smaller $r$ and more refined graphs
and one can study how physical results transform with these
refinements. Third, since the smoothening procedure introduces a
scale ($r$), it may account for the known emergence of new scales
in quantum field theories (e.g. of zero rest mass fields) which
are absent in the classical theory. Finally, the smoothening
procedure also introduces a small and subtle degree of
non-locality which could play an important conceptual role.
\bigskip

{\bf Acknowledgments:} We thank Detlev Buchholz, Chris Fewster,
Stefan Hollands, Thomas Thiemann, and Robert Wald for discussions
during the `Quantum field theory on curved space-times' workshop
at the Erwin Schr\"odinger Institute and for subsequent
correspondence. We are grateful to Klaus Fredenhagen for sharing
with us a number of conceptual and technical insights. This work
was supported in part by the NSF grants PHY-0090091, INT97-22514,
the Albert Einstein Institute and the Eberly research funds of
Penn State.

\end{document}